\documentclass[reprint,floatfix,notitlepage,nofootinbib,twocolumn,superscriptaddress,prb]{revtex4-1}
\usepackage{amsmath}
\usepackage{amssymb}
\usepackage{graphicx}
\usepackage[tight]{subfigure}
\usepackage{enumitem}
\usepackage{soul}
\usepackage{cancel}
\usepackage{tikz}
\usepackage{pifont}
\usepackage{xspace}
\usepackage{comment}

\makeatletter
\renewcommand{\p@subsection}{}
\renewcommand{\p@subsubsection}{}
\makeatother

\usepackage[english]{babel}
\makeatletter
\def\bbl@set@language#1{%
  \edef\languagename{%
    \ifnum\escapechar=\expandafter`\string#1\@empty
    \else\string#1\@empty\fi}%
  \@ifundefined{babel@language@alias@\languagename}{}{%
    \edef\languagename{\@nameuse{babel@language@alias@\languagename}}%
  }%
  \select@language{\languagename}%
  \expandafter\ifx\csname date\languagename\endcsname\relax\else
    \if@filesw
      \protected@write\@auxout{}{\string\select@language{\languagename}}%
      \bbl@for\bbl@tempa\BabelContentsFiles{%
        \addtocontents{\bbl@tempa}{\xstring\select@language{\languagename}}}%
      \bbl@usehooks{write}{}%
    \fi
  \fi}
\newcommand{\DeclareLanguageAlias}[2]{%
  \global\@namedef{babel@language@alias@#1}{#2}%
}
\makeatother
\DeclareLanguageAlias{en}{english}

\usetikzlibrary{arrows}
\usetikzlibrary{intersections}
\usetikzlibrary{shapes.geometric}
\usetikzlibrary{decorations.pathmorphing, patterns,shapes,fixedpointarithmetic}
\usetikzlibrary{decorations.markings}

\tikzset{
  mid arrow/.style={postaction={decorate,decoration={
        markings,
        mark=at position .575 with {\arrow{stealth}}
      }}},
  end arrow/.style={postaction={decorate,decoration={
        markings,
        mark=at position 1 with {\arrow{stealth}}
      }}},
  snake arrow/.style={fixed point arithmetic, decorate, decoration={snake,amplitude=2pt, segment length=11pt},postaction={decoration={markings,mark=at position 0.625 with {\arrow{stealth}}},decorate}},
}

\usepackage{bm}

\usepackage[pdfusetitle,colorlinks=true,citecolor=blue,linkcolor=magenta]{hyperref}

\graphicspath{{./}{./images/}}
\begin{document}
\title{Non-unitary free boson dynamics and the boson sampling problem}

\author{Xiao Chen}
\affiliation{Department of Physics, Boston College, Chestnut Hill, MA 02467, USA}

\begin{abstract}
We explore the free boson unitary dynamics subject to repeated random forced measurement. The input state is chosen as a Fock state in real space with the particle number conserved in the entire dynamics. We show that each boson is performing a non-unitary quantum walk in real space and its dynamics can be mapped to directed polymers in a random medium with complex amplitude. We numerically show that in the one dimensional system, when the measurement strength is finite, the system is in the frozen phase with all the bosons localized in the real space. Furthermore, these single particle wave functions take the same probability distribution in real space after long time evolution. Due to this property, the boson sampling for the output state becomes easy to solve. We further investigate circuit with non-local unitary dynamics and numerically demonstrate that there could exist a phase transition from a localized phase to a delocalized phase by varying the measurement strength. 
 
\end{abstract}

\maketitle

\section{Introduction}
Monitored open quantum systems have attracted a lot of attention in the last few years and have broaden our understanding on non-equilibrium quantum dynamics. It is shown that by varying the monitoring strength, there could exist an entanglement transition from a highly entangled volume law phase to a disentangled area law phase \cite{skinner2019measurement,Li_2018,Chan_2019}. A simple cartoon model to demonstrate this physics is through constructing a hybrid circuit model composed of both unitary gates and local measurement gates. By tuning the measurement rate, both numerical and analytical calculations imply the existence of the phase transition at the level of the quantum trajectories\cite{Li_2018,skinner2019measurement,Li_2019,li2020conformal,gullans2020dynamical,choi2020quantum,bao2020theory,jian2019measurementinduced,nahum2020measurement,bao2021symmetry,Liu_2021,jian2021syk}. 

The observation of entanglement phase transition in open quantum system has broad applications on quantum information research, such as quantum error correction and quantum complexity. Firstly, the non-thermal volume law phase with weak measurement can be interpreted as a dynamically generated quantum error correction code in which the quantum information is protected by the unitary dynamics and is resilient to local errors caused by weak measurement\cite{choi2020quantum,gullans2020dynamical,li2020statistical,fan2020self}. Secondly, the entanglement transition naturally implies a computational complexity transition. In the highly entangled volume law phase, storing the wave function in general requires exponential resource and therefore its dynamics is difficult to simulate on the classical computer. On the other hand, in the disentangled area law phase, the wave function can be efficiently represented by a matrix product state with small bond dimension and is simulable on the classical computer\cite{Schuch_2008}. Furthermore, it has implication on the quantum sampling problem which samples the probability distribution in a quantum system upon measurement in some basis. Many of these sampling problems such as instantaneous quantum computation (IQP)\cite{Shepherd_2009,Bremner_2010}
 and random circuit sampling problem\cite{Bouland_2018,Boixo_2018,google} can go through some complexity transitions which can be understood in terms of the entanglement transition\cite{vijay2020measurementdriven,napp2020efficient}. One interesting example is the two dimensional (2d) random shallow circuit, which undergoes a sampling complexity transition by increasing the circuit depth $D$\cite{terhal2002adaptive}. This transition can be effectively mapped to a 1+1d hybrid circuit with one spatial dimension in the original 2d circuit treated as the time direction and $1/D$ related to the measurement strength\cite{napp2020efficient}.
 
 Motivated by these works on quantum complexity transition, in this paper, we are going to consider the boson sampling problem subject to random forced measurement. For the non-interacting bosons undergoing the unitary dynamics, the output distribution is described by matrix permanent which in general is hard to simulate on the classical computer\cite{aaronson2010computational}. This simulation can run on the quantum computer and can be used as a benchmark test for quantum computational supremacy\cite{Zhong_2020}. In this paper, we will show that once the dynamics become non-unitary, the sampling is an easy task on the classical computer.

\begin{figure}
    \centering
     \includegraphics[scale=0.45]{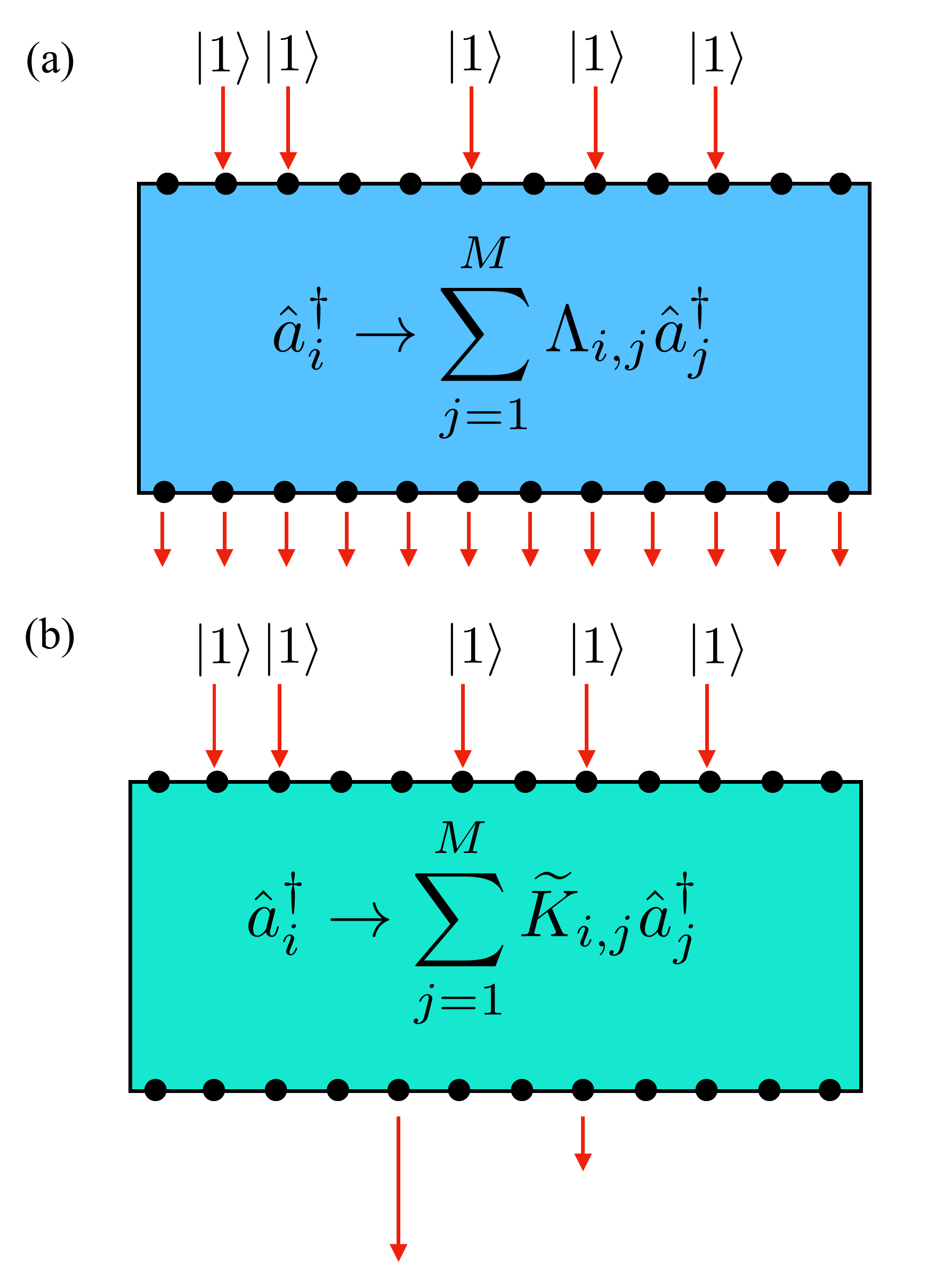}
    \caption{(a) The boson sampling problem in unitary dynamics. The input state is a Fock state in real space. Under the unitary time evolution, the single particle wave function spreads out and becomes extended in the real space. In general, for the output many-body state, the boson density has a near uniform distribution and sampling in the Fock space is a hard task on the classical computer. The length of the red arrow denotes the particle density. (b) The boson sampling in the non-unitary dynamics. The input is the same as in (a) and the dynamics can now be described by DPRM with complex amplitude. For the output state, the bosons are localized at a few sites. The location of the bosons (location for the red arrow) fluctuate with the time and can be characterized by a large wandering exponent $\xi=2/3$. Due to the localization property of the bosons, the output sampling problem is simulable on the classical computer.}
    \label{fig:boson_cartoon}
\end{figure}

Another motivation to work on this problem is based on the recent works on non-unitary dynamics in free systems. 
In the free fermion system, it is known that under the pure unitary dynamics, the steady state can obey volume law entanglement scaling caused by the spreading of quasi-particle pairs\cite{Calabrese_2005}. However, different from the interacting system, this state is unstable in the presence of the weak measurement\cite{Cao_2019,Chan_2019}. A small but finite weak measurement will destroy the volume law scaling and give rise to a critical phase with logarithmic entanglement scaling\cite{Chen_2020,alberton2020trajectory}. The instability of volume law phase also occurs in the free boson dynamics. In Ref.~\onlinecite{zhou2021nonunitary}, it is shown that for a Gaussian bosonic circuit subject to Gaussian measurement, the steady state entanglement entropy obeys area law scaling due to the disparity of the competing time scales to entangle and
disentangle the degrees of freedom. In this paper, we consider a slightly different setup for free boson dynamics. We impose particle conservation law and study the dynamics starting from a Fock state in real space which is not a Gaussian state anymore (See Fig.~\ref{fig:boson_cartoon}). Since this is a free boson dynamics, the quantum state at any time can be represented as a product of boson creation operators acting on the vacuum. We demonstrate that each particle is undergoing non-unitary quantum walk which can be effectively mapped to directed polymers in a random medium (DPRM) with complex amplitude\cite{Huse_Henley,Cook_Derrida,Zhang_1989}. Recently, many entanglement dynamics problems are found to be related to DPRM\cite{Nahum_2017,zhou2019emergent,li2021entanglement}. In this paper, we focus on the one dimensional system and numerically show that all of these boson creation operators become the same and are localized in space at a few sites after long time evolution (See Fig.~\ref{fig:boson_cartoon} (b)). The location of the particles have strong fluctuation in spatial direction and can be characterized by the wandering exponent $\xi=2/3$, the same as that in the standard DPRM problem with positive weight\cite{Huse_Henley,Kardar_Nelson,Huse_Henley_Fisher,Kardar_Zhang,Kardar_1987}. Therefore sampling the output distribution in the Fock space becomes simulable on the classical computer. We further generalize the above analysis to higher dimension and consider a simple model where the unitary dynamics has no locality. We show that in this system, there could be a transition from localized phase to delocalized phase. In the latter, the particles are extended in real space and the sampling becomes intractable classically. 

The rest of the paper is organized as follows. In Sec.~\ref{sec:review}, we review the boson sampling problem. In Sec.~\ref{sec:non_unitary}, we discuss the non-unitary free boson dynamics and establish the connection with DPRM problems. In Sec.~\ref{sec:numerical}, we perform numerical simulation for our model and study the localized phase in it. In Sec.~\ref{sec:conclusion}, we summarize our results and discuss several
interesting problems for future work. 

\section{Review of Boson sampling problem}
\label{sec:review}
In the boson sampling problem,  we prepare the input state as a Fock state and sample the output probability in Fock basis. Consider a $N$ bosons Fock state $|\psi_0\rangle=\hat a_{i_1}^\dag \hat a_{i_2}^\dag\cdots  \hat a_{i_n}^\dag\cdots \hat a_{i_N}^\dag |0\rangle$ defined on a lattice of $M$ sites, where $\hat a_{i_n}^\dag$ denotes the boson creation operator on site $i_n$. Under particle-conserving unitary time evolution, the bosons will hop on the lattice and the output many-body state becomes a superposition of Fock states, i.e., \begin{equation}
    |\psi\rangle=\sum_{C}\gamma_C|n_1^C,n_2^C,\cdots,n_M^C\rangle,
\end{equation}
where $n_i^C$ is the number of boson in the $i$th site in the  configuration $C$ and $|\gamma_C|^2$ is the corresponding probability for this configuration. In the non-interacting case, the output state is given by
\begin{align}
|\psi\rangle=\hat b_{1}^\dag \hat b_{2}^\dag\cdots  \hat b_{i}^\dag\cdots \hat b_{N}^\dag |0\rangle,
\end{align}
where $\hat b^
\dag_i$ is a Heisenberg operator associated with $\hat a_i^\dag$ and is simply given by the linear transformation on $\hat a_i^\dag$,
\begin{equation}
    \hat b^\dag_i\equiv\hat{U} \hat a_i^\dag \hat{U}^\dag=\sum_{j=1}^M\Lambda_{i,j}\hat a_j^\dag,
    \label{eq:u_evo}
\end{equation}
where $\hat{U}$ is the unitary evolution operator and $\Lambda$ is the associated $M\times M$ unitary matrix defined on the boson creation operators.

Using this result, it is easy to show that the amplitude of Fock basis for the output state is
\begin{equation}
    \gamma_C=\frac{\mbox{Per}(\Lambda_C)}{\sqrt{n_1^C!\cdots n_M^C!}},
\end{equation}
where $\Lambda_C$ is an $N\times N$ submatrix of $\Lambda$ and $\mbox{Per}(\Lambda_C)$ is the permanent of $\Lambda_C$. Since computing the permanent of complex matrices is $\#P$-Complete, simulating boson sampling on classical computer requires exponential resource and in general can only be done on small systems on classical computer\cite{aaronson2010computational}. 

The non-interacting unitary evolution can be generated by a quantum circuit with two-site gates or quadratic Hermitian Hamiltonians. For the quantum circuit or the Hamiltonian with a {\it local} structure, at early time, there is a light cone structure for the spreading of boson created by $\hat b^\dag_i$. The boson can spread out ballistically or diffusively depending on whether or not if there is randomness in the dynamics. In both cases, if the initial state has $N\ll M$ and bosons are well separated from each other, at short time, $\hat b^\dag_i$ is localized in a small regime. As a consequence, the bosons are distinguishable and boson sampling is classically easy to solve \cite{Brod_2015,Deshpande_2018,maskara2020complexity}. As time evolves, $\hat b^\dag_i$ becomes highly non-local and the quantum interference between bosons is strong. Therefore the sampling problem is intractable classically. This indicates that under time evolution, there exists a dynamical transition from easy to hard in terms of sampling complexity \cite{Deshpande_2018}. Below we are going to consider the non-unitary free boson dynamics and we will show that the sampling problem becomes easy to solve.



\section{Non-unitary dynamics}
\label{sec:non_unitary}

Here we consider the following free boson non-unitary random dynamics described by
\begin{align}
    \hat U=\prod_{t=1}^T \hat U_I (t) \hat U_R (t),
\end{align}
which consists of both unitary and imaginary evolutions. In each period, $\hat U_R(t)$ denotes the unitary evolution operator at time $t$ and can be generated by local two-site gates or local Hamiltonian with the detail explained later. $\hat U_I(t)\equiv \exp(-2\beta \hat H_2(t))$ is random imaginary time evolution operator, where $\beta\geq 0$ denotes the strength of the imaginary evolution and the Hamiltonian $\hat H_2(t)$ is a simple random onsite potential
\begin{align}
    \hat H_2(t)=\sum_x \lambda_{x,t} \hat a_x^\dag \hat a_x,
    \label{eq:imag}
\end{align}
where $\lambda_{x,t}$ is random in both space and time. For simplicity, in the numerical simulation, we consider $\lambda_{x,t}$ takes a uniform distribution between 0 and 1. In $\hat U_I(t)$, each term $\exp(-2\beta \lambda_{x,t}\hat a_x^\dag \hat a_x)$ can be treated as a forced measurement on the particle number at $x$th site.

We are interested in the pure wave function dynamics,
\begin{align}
    |\psi(T)\rangle=\frac{\hat U(T)}{\sqrt{Z}}|\psi_0\rangle
\end{align}
where $Z=\langle \psi_0 |\hat U^
\dag(T) U (T)|\psi_0\rangle$ is the normalization factor and $|\psi_0\rangle$ is a Fock state generated by $N$ boson creation operators. This wave function can be rewritten as 
\begin{align}
    |\psi(T)\rangle\sim \hat U \hat a_1^\dag \hat U^{-1}\hat U \hat a_2^\dag \hat U^{-1}\cdots \hat U \hat a_N^\dag \hat U^{-1} \hat U|0\rangle.
\end{align}
Since this is free boson dynamics, $\hat U \hat a_i^\dag \hat U^{-1}$ is proportional to a boson creation operator. As shown in Eq.\eqref{eq:u_evo}, under unitary evolution, $\hat a^\dag_i$ is transformed by an associated unitary matrix $\Lambda^R$. Under the imaginary time evolution, it is transformed in a similar way, except now the transformation matrix is non-unitary, i.e., 
\begin{align}
    \hat U_I \hat a_i^\dag \hat U^{-1}_I=\sum_{j=1}^M \Lambda^I_{i,j}\hat a_j^\dag.
\end{align}
In particular, if we take $\hat U_I$ to be generated by $\hat H_2$, $\Lambda^I$ is a diagonal matrix with $\Lambda^I_{ii}=\exp(-2\beta \lambda_i)$. Combining these together, we have
\begin{align}
    |\psi(T)\rangle=\hat b_1^\dag \hat b_2^\dag\cdots \hat b_N^\dag|0\rangle
    \label{eq:wf_time}
\end{align}
where $\hat b_i^\dag=\sum_{j=1}^M \widetilde{K}_{i,j}\hat a_j^\dag$ is still a linear combination of $\hat a_j^\dag$.  Here we have $M\times M$ matrix $K=\prod_{t=1}^T\Lambda_t^I\Lambda_t^R$ and the normalized $\widetilde{K}_{i,j}\equiv K_{i,j}/\sqrt{\sum_{j}|K_{i,j}|^2}$.

\begin{figure}
    \centering
     \includegraphics[scale=0.45]{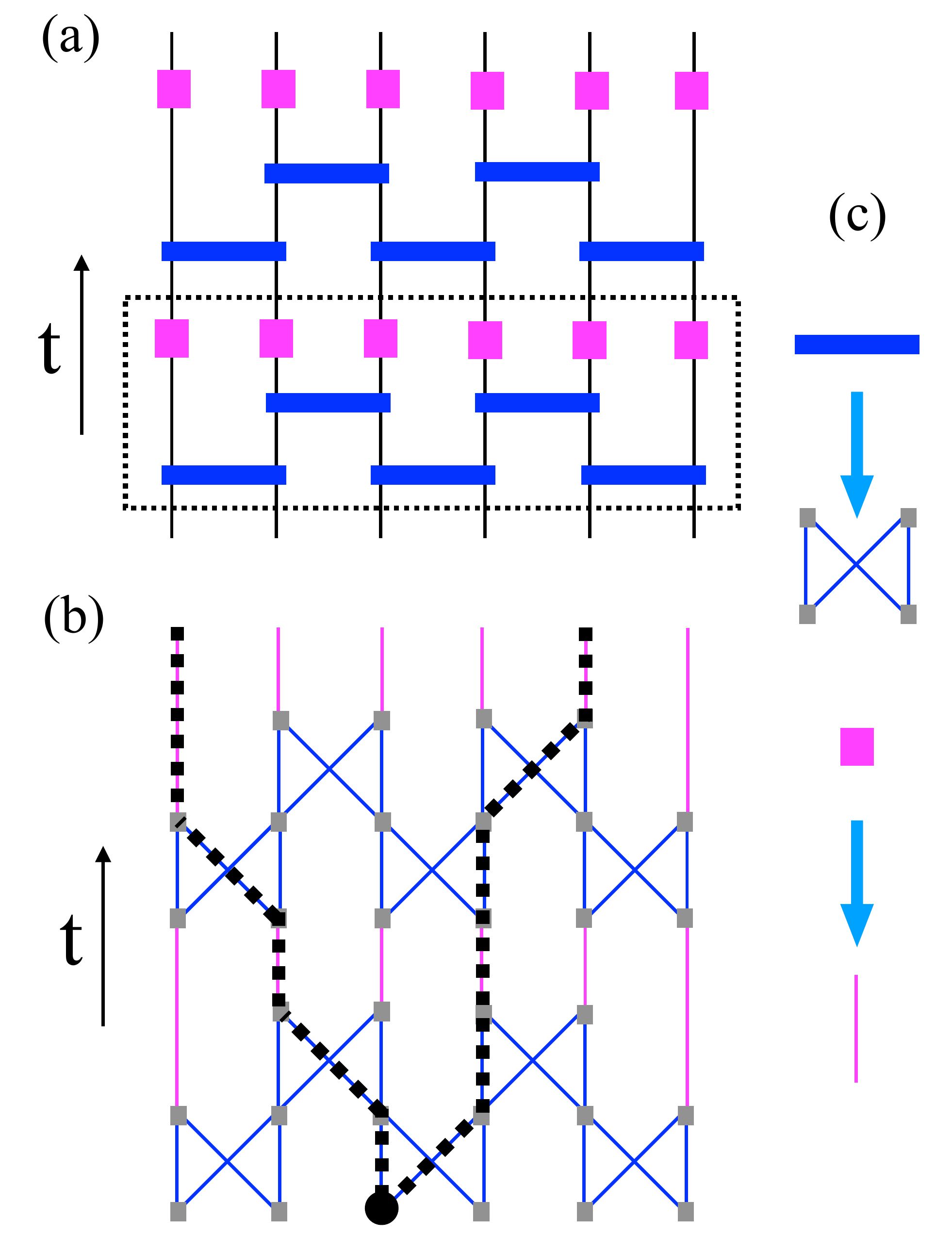}
    \caption{(a) The cartoon for the non-unitary circuit composed of both unitary two-site gates and imaginary one-site gates. The unitary part has a brick wall structure and the imaginary evolution is determined by Eq.\eqref{eq:imag}. The dashed box denotes time evolution in one time step. (b) The corresponding lattice for the directed polymer problem. Here we draw two possible paths for the directed polymer. The weight for the blue bond is determined by the unitary evolution while the weight for the pink bond is determined by the imaginary evolution. (c) describes the rule of generating the lattice in (b) from the circuit in (a).}
    \label{fig:cartoon}
\end{figure}

The calculation of the coefficient $K_{i,j}$ in $\hat b_i^\dag$ has an interesting connection with DPRM\cite{Huse_Henley}. For simplicity, we first limit our analyze to the circuit with spatial dimension $d=1$ and take the unitary evolution generated by the two-site gates (see Fig.~\ref{fig:cartoon} (a)). Each unitary gate has an associated $2\times 2$ unitary matrix $A$ acting on the boson creation operator. Its four matrix elements are labelled as $A^a$ with $a=1,2,3,4$. Therefore we have
\begin{align}
    K_{i,j}=\sum_{P}\prod_t e^{-2\beta \lambda_{x,t}}A_{t}^{1,a}A_{t}^{2,a^\prime},
    \label{eq:K_ij}
\end{align}
where the superscript $1$ and $2$ in $A$ describe the first and the second layers of the unitary gates respectively at time $t$ and $\exp(-2\beta \lambda_{x,t})$ is contributed by the imaginary evolution. 
Combining the unitary and imaginary evolutions in this way, $K_{i,j}$ is the amplitude for all the oriented paths $\{P\}$ starting from $i$th site at $t=0$ and ending at $j$th site at $t=T$ on the lattice  (See Fig.~\ref{fig:cartoon} (b)). Furthermore, $|\widetilde{K}_{i,j}|^2$ is the probability for the endpoint at $j$th site. Notice that different from the standard DPRM, both $A^a$ and $K_{i,j}$ can take complex value \cite{Cook_Derrida,Zhang_1989}. 

In the standard 1d DPRM, $K_{i,j}$ is sum of the weight for paths starting from $i$th site at $t=0$ and ending at $j$th site at $t=T$. For simplicity, we can set $A^a$ to be a positive constant and  the random medium is caused by $\lambda_{x,t}$, which is defined at the pink bond in Fig.~\ref{fig:cartoon} (b). In this DPRM problem, it is known that when $\beta$ is finite, the randomness in $\lambda_{x,t}$ is always relevant and drives the system into a disorder dominant phase. The endpoints of the paths have strong fluctuation and can be characterized by the large wandering exponent $\xi=2/3$ defined in $\langle X^2\rangle\sim T^{2\xi}$ \cite{Huse_Henley,Kardar_Nelson,Huse_Henley_Fisher,Kardar_Zhang,Kardar_1987}. Another feature of this phase is that the polymer is frozen in a few paths and this further leads to the localization of the endpoint at time $T$\cite{carmona2006strong,comets2003directed}.

\section{Numerical simulation}
\label{sec:numerical}
\begin{figure}
    \centering
     \includegraphics[scale=0.45]{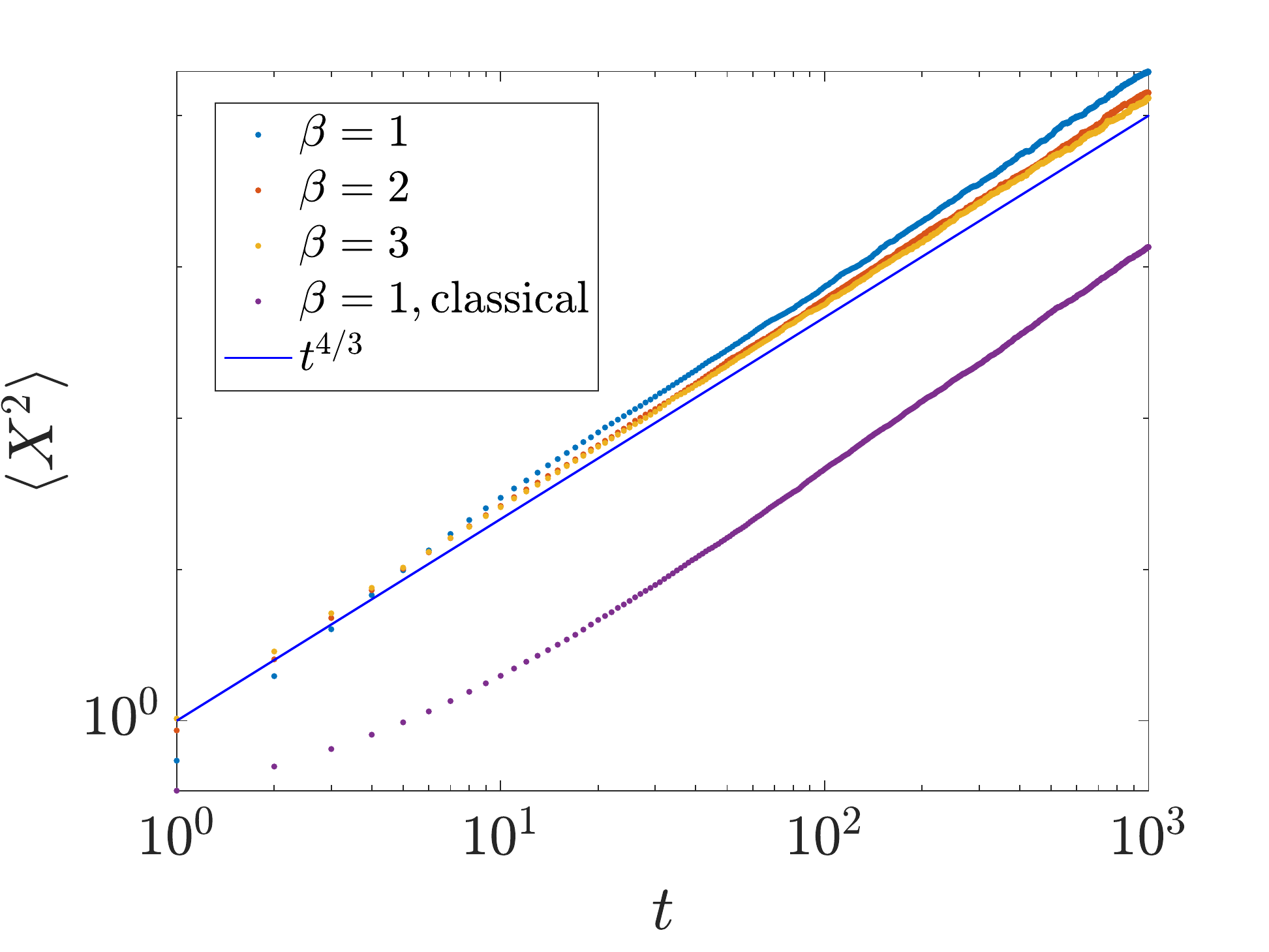}
    \caption{The fluctuation of $\langle X^2\rangle$ vs time. In the numerical simulation, $K_{i,j}$ is evaluated by using Eq.\eqref{eq:K_ij}. In both the quantum model (with $A$ defined in Eq.\eqref{eq:matrix_A}) and classical model (with $A^a=1$), we observe that $\langle X^2\rangle\sim t^{4/3}$. } 
    \label{fig:endpoint_var}
\end{figure}

In our model, $A^a$ can be a complex number. It is interesting to see if such a difference can lead to a qualitatively different behavior. Below we numerically simulate the random dynamics and we choose \begin{align}
    A=\frac{1}{\sqrt{2}}\begin{pmatrix}
    1 & -i \\ -i & 1
    \end{pmatrix}.
    \label{eq:matrix_A}
\end{align}
We compute the fluctuation $\langle X(t)^2\rangle\equiv\overline{\left[\sum_j j|\widetilde{K}_{i,j}|^2-i\right]^2}$ with $\overline{\bullet}$ denoting average over different circuit realizations. As shown in Fig.~\ref{fig:endpoint_var}, at finite $\beta$, $\langle X(t)^2\rangle\sim t^{4/3}$, the same as the standard DPRM with positive $A^a$. In spite of  possible quantum interference caused by the complex amplitude, it seems that our model still belongs to the standard DPRM universality class.

\begin{figure}
    \centering
     \includegraphics[scale=0.45]{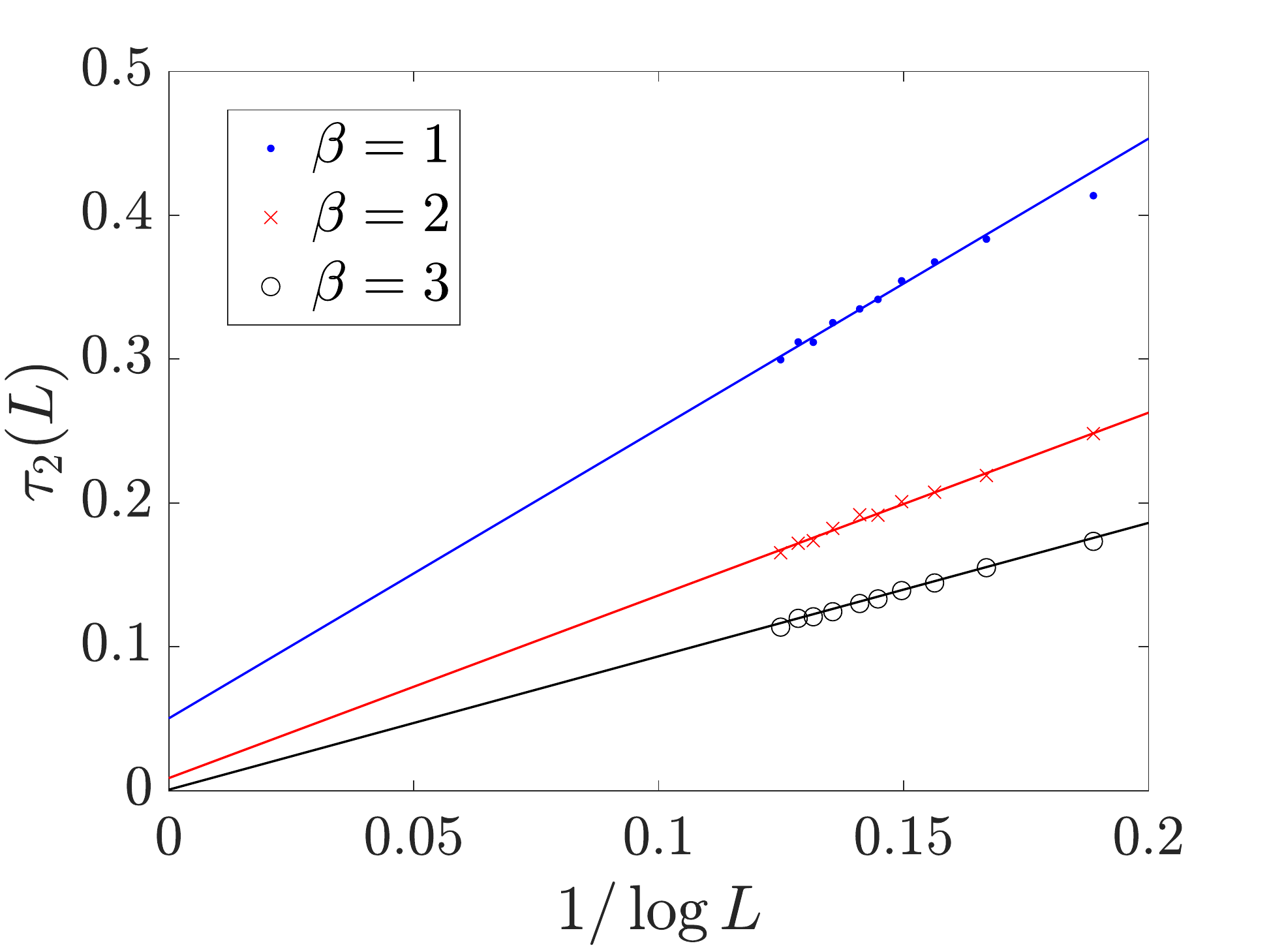}
    \caption{$\tau_2(L)$ for single boson wave function vs $1/\log L$ in the non-unitary circuit described in Fig.~\ref{fig:cartoon} (a).} 
    \label{fig:IPR_local}
\end{figure}

We further look at the endpoint probability distribution $|\widetilde{K}_{i,j}|^2$. In order to quantify how extended the distribution is, we compute its inverse participation ratio (IPR) $I_q=\sum_j|\widetilde{K}_{ij}|^{2q}$ as a function of the system $L$. In this paper we take $q=2$ and we expect that $I_2=a L^{-\tau_2}$. In the limit $L\to\infty$, if the boson creation operator $\hat b_i^\dag$ is localized around a few site, we have $\tau_2=0$. On the other hand, if $\hat b_i^\dag$ is extensive in the space, $\tau_2=1$. In the numerical simulation, we 
define $\tau_2(L)$ at finite $L$ as $-\overline{\log I_2}/\log L$ and we have
\begin{align}
    \tau_2(L)=-\frac{\log a}{\log L}+\tau_2.
\end{align}
 The numerical result for different $L$ is presented in  Fig.~\ref{fig:IPR_local} and we observe that $\tau_2(L)$ decreases as we increase the system size. The extrapolation in the thermodynamic limit $L\to\infty$ is $\tau_2$ which is close to zero at finite $\beta$, indicating that the bosons are localized in real space. We have also numerically checked $\tau_2$ for smaller $\beta<1$, and we notice that although it is nonzero, $\tau_2(L)$ is always decreasing as we increase the system size. We believe that the nonzero $\tau_2$ is due to large finite size at small $\beta$.

\begin{figure}
    \centering
     \includegraphics[scale=0.45]{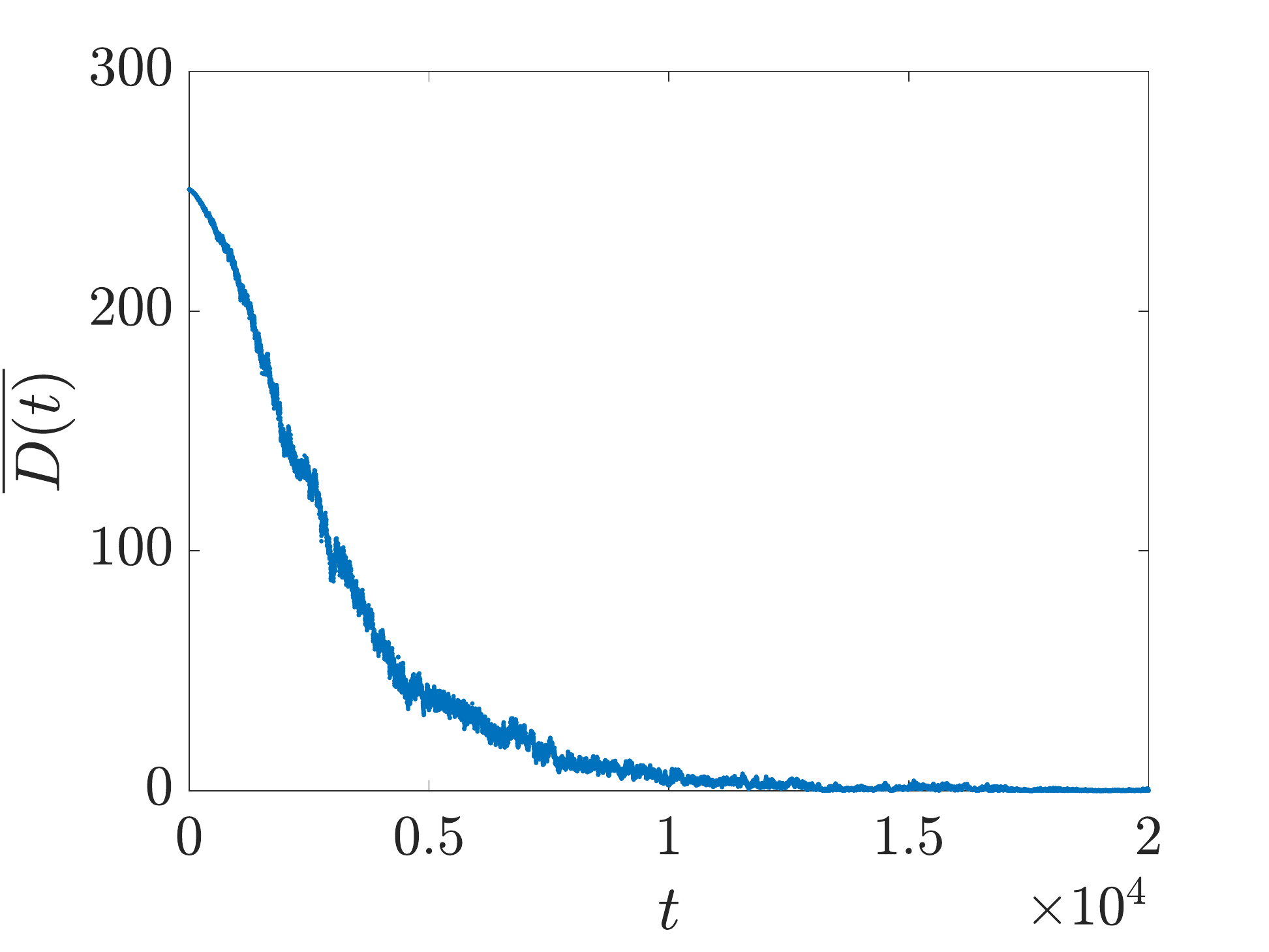}
    \caption{The averaged distance between all pairs of bosons in the non-unitary circuit described in Fig.~\ref{fig:cartoon} (a). In the numerical calculation, $\overline{D(t)}$ is further averaged over $100$ circuit realizations. The system parameters are $L=1000$ and $\beta=2$. The initial state is chosen as a random Fock state with particle number $N=500$. } 
    \label{fig:D}
\end{figure}
For the input many-body state with $N$ bosons, under non-unitary evolution, all of them will be localized. Furthermore, we observe that after long time evolution, all of $\hat b_i^\dag$ approach the same form and can be approximated as 
\begin{align}
    \hat b^\dag\approx\sum_{s=1}^S \alpha_{s}\hat a^\dag_{m_s},
\end{align}
where $\alpha_{s}$ takes nonzero value at $m_s$th site and the total number of these sites $s$ is finite.  This property can be verified by computing the distance $D=|\langle X_i\rangle-\langle X_j\rangle|$ between every pairs of bosons in the system. As shown in Fig.~\ref{fig:D}, the distance decays to zero after long time evolution. We further check the explicit form of $\hat b_i^
\dag$ and find that all of them are the same up to an unimportant overall phase. Therefore, for the output state $|\psi\rangle=(\hat b^\dag)^k|0\rangle$, it can be expanded in the Fock basis as
\begin{align}
|\psi\rangle=\sum_{n_i}\frac{N!}{n_1!\cdots n_S!}\alpha_1^{n_1}\cdots \alpha_S^{n_S}(\hat a_{m_1}^\dag)^{n_1}\cdots (\hat a_{m_S}^\dag)^{n_S}|0\rangle,
\end{align}
where $n_s$ is the number of boson at $m_s$th site and satisfies $n_1+n_2+\cdots+n_S=N$. Since the coefficient for each Fock state is easy to compute, the boson sampling problem is simulable classically. 

\begin{figure}
    \centering
    \subfigure[]{\label{fig:endpoint_var_1} \includegraphics[width=.9\columnwidth]{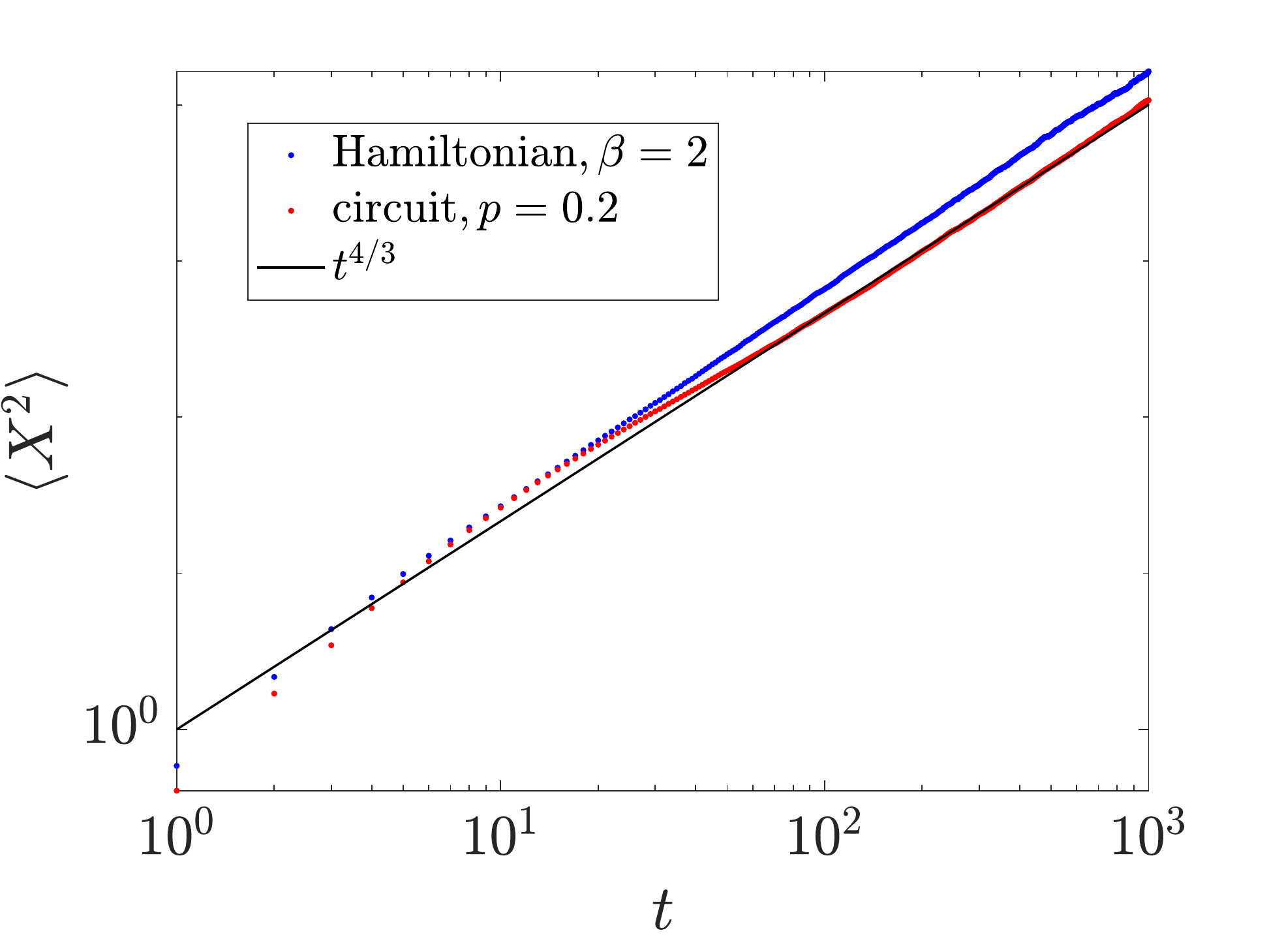}}
    \subfigure[]{\label{fig:IPR_local_1} \includegraphics[width=.9\columnwidth]{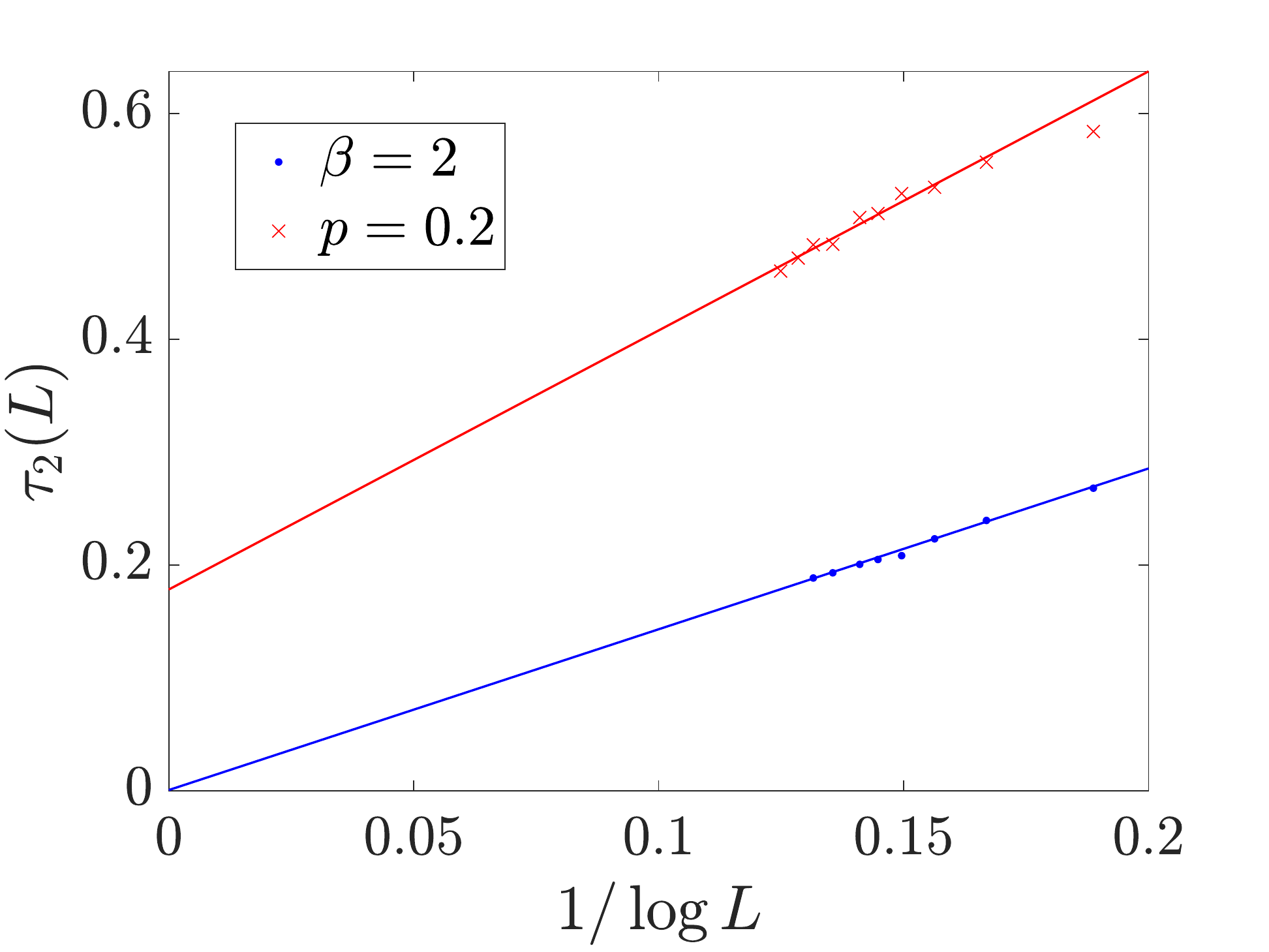}}
    \caption{The fluctuation $\langle X^2\rangle$ (a) and $\tau_2(L)$ (b) in various non-unitary dynamics. For the blue dot/curve, the unitary dynamics is governed by Hamiltonian described in Eq.\eqref{eq:local_H}. For the red dot/curve, the unitary dynamics is generated by the quantum circuit and the imaginary evolution is described by the local forced measurement projecting to the vacuum state.}
    \label{fig:Local_1}
\end{figure}

In the above analysis, the unitary dynamics is generated by two-site gates in Fig.~\ref{fig:cartoon} (a). We could replace the two-layer unitary circuit at every time step by a quadratic Hamiltonian with local hopping terms
\begin{align}
    H=\sum_{i}\hat b^\dag_i\hat b_{i+1}+h.c..
    \label{eq:local_H}
\end{align}
Combining this with random imaginary evolution, we observe that $\langle X(t)^2\rangle\sim t^{4/3}$ and the bosons are still localized at a few sites (See Fig.~\ref{fig:Local_1}). We further replace the random imaginary evolution by local measurement projected to vacuum. At every time step, after the unitary evolution, we apply this forced local measurement randomly with rate $p$ and we observe the same localization phenomenon as shown in Fig.~\ref{fig:Local_1}.

\begin{figure}
    \centering
     \includegraphics[scale=0.45]{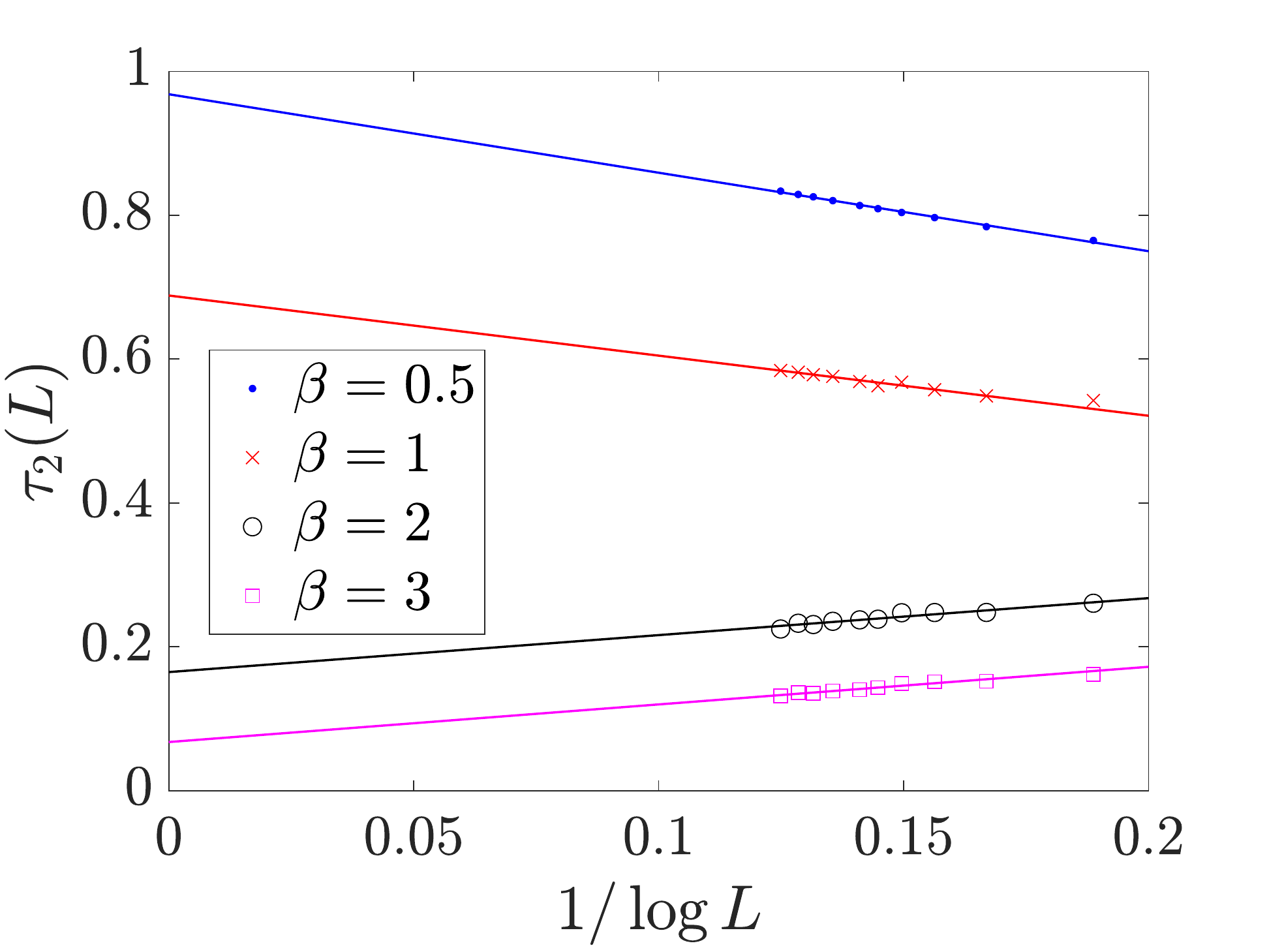}
    \caption{$\tau_2(L)$ for single boson wave function vs $1/\log L$ in the non-local non-unitary dynamics.} 
    \label{fig:IPR_all}
\end{figure}

For the above 1+1d non-unitary circuits, we observe that the dynamics is described by the standard DPRM in spite of the complex amplitude. At finite $\beta$ or finite $p$, there only exists a frozen phase dominated by the disorder. This may not be true when the spatial dimension is large or the unitary dynamics is non-local. Below we construct a simple circuit without locality: at every time step, we randomly pick up $N/2$ pairs of sites and apply the two-site unitary gates described by Eq.\eqref{eq:matrix_A}. For such a unitary dynamics subject to repeated random imaginary evolution, we observe a phase transition from localized phase to delocalized phase by tuning $\beta$, which further implies an easy to hard transition in terms of sampling complexity. As shown in Fig.~\ref{fig:IPR_all}, when $\beta$ is large, $\tau_2(L)$ decreases with $\beta$ and approaches zero in the limit $L\to\infty$, the same as the 1d case. On the other hand, when $\beta$ is small, $\tau_2(L)$ increases when we increase $L$, indicating that the wave function is delocalized. Numerically we observe that the transition occurs when $1<\beta <2$. When $\beta=0.5$,   $\tau_2(L\to\infty)$ is very close to 1, implying that the imaginary evolution is irrelevant and the single particle wave function is fully extended, the same as the pure unitary dynamics. When $\beta=1$, $\tau_2(L\to\infty)$ is significantly smaller than 1 and it is unclear if the large deviation from 1 is a finite size effect or not. It is interesting to examine if there exists an intermediate phase between the localized phase and the fully extended phase. We leave the detailed study of this for the future.

\section{Conclusion}
\label{sec:conclusion}
In this paper, we explore the non-unitary free boson dynamics. We show that the bosons are performing non-unitary quantum walk on the lattice and the dynamics can be described by DPRM with complex amplitude. For the system living on the one dimensional lattice, at finite $\beta$ (or finite measurement rate $p$), after long time evolution, all of the bosons are localized on a few sites, making the boson sampling problem simulable on the classical computer. We further consider a system in which the unitary dynamics does not have locality and we show that there exists a delocalization phase transition by tuning $\beta$. 

The numerical simulation in 1d system indicates that at finite $\beta$, our DPRM model with complex amplitude exhibits the same frozen phase as described in the standard DPRM model with positive weight. For the standard DPRM, when $d=1,2$, at finite $\beta$, there only exists the frozen phase dominated by the disorder. When $d\geq 3$, by decreasing $\beta$, the polymer can undergo a phase transition from the frozen phase to the free phase where the disorder is irrelevant\cite{Imbrie_1988}. It would be interesting to study our DPRM models in high dimensions and investigate the possible phase transitions in them and check if there exists other novel quantum phases beyond the localized and extended phases.  

In the 1d random non-unitary free fermion dynamics, there exists a critical steady state $|\psi\rangle=\prod_i \hat f_i^\dag|0\rangle$ where $\{f_i,f_j \}=0$ when $i\neq j$ and the single fermion wave function created by $\hat f_i^\dag$ exhibits multifractal behavior\cite{iaconis2021multifractality}. Such critical state is absent in the bosonic case, at least in our current setup. This difference is due to the Pauli exclusion principle in the fermionic system, where there could only be at most one fermion at each site. At the operational level, these $N$ fermion creation operators $\{\hat f_i^\dag\}$ are vectors with length $L$ which form the orthonormal basis for the $N$ vectors associated with $\{\hat b_i^\dag\}$. Such a difference leads to a remarkably different behaviors between the free boson and free fermion wave functions. It would be interesting to better understand the mathematics behind this difference in the non-unitary dynamics in the future.

The non-unitary dynamics in our model involves the local forced measurement. We could also consider a non-unitary circuit subject to the projective measurement. It would be interesting to see if this can lead to different dynamical phase transition and examine potential sampling complexity transition in it. In addition, we could also use some known results in the non-unitary dynamics to study other quantum complexity problems. We leave all these interesting questions for the future study.

\acknowledgements
XC thanks Qicheng Tang, Wei Zhu and Tianci Zhou for collaborations on
related topics.

\bibliography{ref}

\end{document}